\documentclass[aps,prc,preprint,tightenlines,floatfix,groupedaddress,
nofootinbib,showpacs,preprintnumbers,amsmath,amssymb,superscriptaddress]
{revtex4}
\usepackage{graphicx}
\usepackage{dcolumn}
\usepackage{mathrsfs}
\usepackage{bm}
\begin{document}

\title{Nuclear Physics of Neutron Stars }
\author{J. Piekarewicz\footnote{\tt e-mail: jpiekarewicz@fsu.edu}}
\affiliation{Department of Physics, Florida State University,
             Tallahassee, FL {\sl 32306}}
\date{\today}

\begin{abstract}
Understanding the equation of state (EOS) of cold nuclear matter,
namely, the relation between the pressure and energy density, is 
a central goal of nuclear physics that cuts across a variety of
disciplines. Indeed, the limits of nuclear existence, the collision 
of heavy ions, the structure of neutron stars, and the dynamics of
core-collapse supernova, all depend critically on the equation of
state of hadronic matter. In this contribution I will concentrate on
the special role that nuclear physics plays in constraining the EOS of 
cold baryonic matter and its impact on the properties of neutron stars. 
\end{abstract}
\maketitle

\section{Introduction}

A neutron star is a gold mine for the study of the phase diagram of
cold baryonic matter. Indeed, a remarkable fact about spherically
symmetric neutron stars in hydrostatic equilibrium --- the so-called 
{\sl Schwarzschild stars} --- is that the only physics that they are 
sensitive to is the equation of state of cold, neutron-rich matter. 
As such, neutron stars provide a myriad of observables that may be 
used to constrain poorly known aspects of the nuclear interaction 
under extreme conditions ({\sl both high and low}) of density. 
Contrary to the most common perception of a neutron star as a uniform
assembly of neutrons packed to densities that may exceed that of
normal nuclei by up to an order of magnitude, the reality is far
different and significantly more interesting.  For example, the mere
{\sl model-independent} fact that hydrostatic equilibrium must be
maintained throughout the star, generates a range of densities that
span over 11 orders of magnitude; from about $10^{4}$ to 
$10^{15}~{\rm g/cm^{3}}$ (nuclear-matter saturation density equals
$\rho_{0}\!=\!2.48\times 10^{14}{\rm g/cm^{3}}$). What novels phases 
of baryonic matter emerge under these extreme conditions is at
present both fascinating and unknown. As we shall see, most of the 
exotic phases predicted to exist in neutron stars can not be realized 
under normal laboratory conditions. Whereas most of these phases have 
a fleeting existence in the laboratory, they become stable in neutron 
stars as a consequence of the presence of its enormous gravitational 
field. Understanding the physics underlying these phases will require a
concerted and sustained effort from laboratory experiments, astronomical 
observations, and theoretical analyses. In this contribution I will aim 
to establish the particular role that nuclear physics plays in
establishing some robust and enlightening correlations between the 
properties of finite nuclei and neutron stars.

\section{Tomography of a Neutron Star}

\begin{figure}[h]
  \includegraphics[height=2.7in]{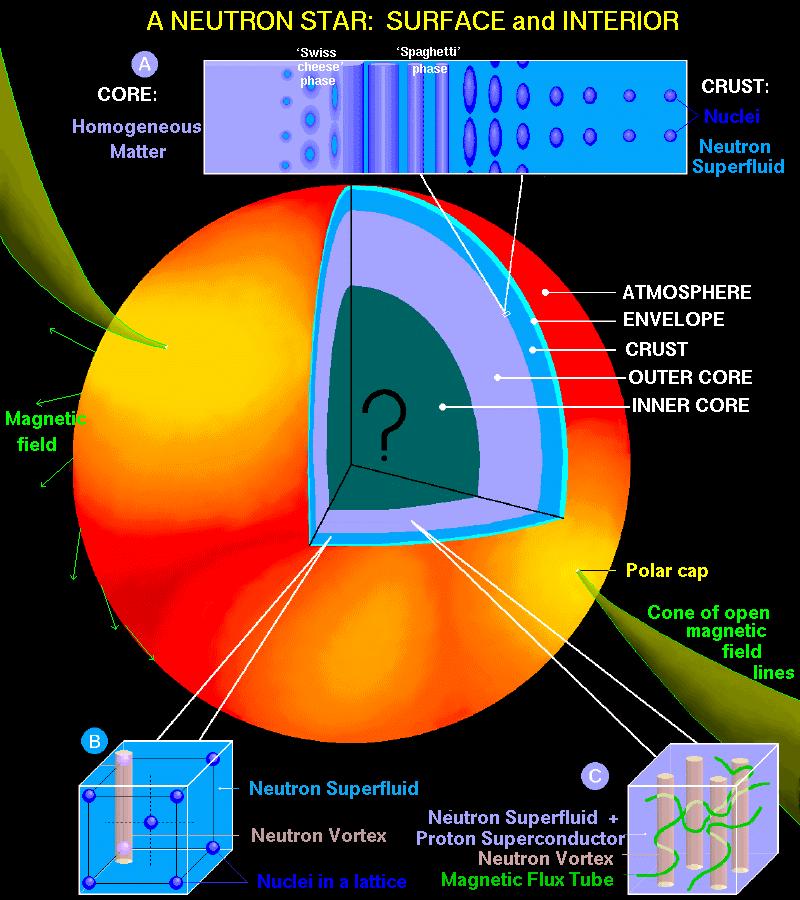}
 \vspace{-0.2cm}
 \caption{A rendition of the structure and phases of
          a neutron star (courtesy of Dany Page).}
 \label{Fig1}
\end{figure}

Neutron stars contain a non-uniform crust above a uniform liquid
mantle that, in turn, is located above a possible exotic core. Figure~\ref{Fig1} displays what is believed to be an accurate 
rendition of a neutron star (courtesy of Dany Page).

\subsection{The Outer Crust}
\label{OuterCrust}

Neutron stars contain a non-uniform crust above the uniform liquid
mantle (or outer core).  The solid outer crust is understood as the
region of the star spanning about 7 orders of magnitude in density;
from about $10^{4}{\rm g/cm^{3}}$ to $4\times 10^{11}{\rm
g/cm^{3}}$~\cite{Baym:1971pw}. At these densities, the electrons ---
which are an essential component of the star in order to maintain
charge neutrality --- have been pressure ionized and move freely
throughout the crust.  In contrast, at these extremely low {\sl
nuclear} densities the uniform state is unstable against cluster
formation. Hence, at the lowest densities of the outer crust, the
nucleons cluster into ${}^{56}$Fe nuclei which in turn arrange
themselves in a crystalline face-centered-cubic lattice in order to
minimize their overall Coulomb repulsion.  However, as the density
increases (as one moves deeper into the star) ${}^{56}$Fe --- with the
smallest mass per baryon --- is no longer the most energetically
favorable nucleus. This is because the electronic contribution to the
energy increases faster with density than the corresponding nuclear
contribution~\cite{Baym:1971pw,RocaMaza:2008ja}.  As a result, it
becomes energetically advantageous for the energetic electrons to
capture on the protons and for the excess energy to be carried away by
neutrinos. The resulting nuclear lattice is now made of nuclei having
a neutron excess larger than that of ${}^{56}$Fe.  As the density
continues to increase, the nuclear system evolves into a Coulomb
lattice of progressively more neutron-rich nuclei until a {\sl
``critical''} density of about $4\times 10^{11}{\rm g/cm^{3}}$ is
reached. At this point the nuclei are unable to hold any more
neutrons: {\sl the neutron drip line has been reached}.  Thus, the
physics of the outer crust is dominated by an interesting competition
between an electronic contribution that attempts to drive the system
toward more neutron rich nuclei ({\sl i.e.,} smaller proton fractions)
and a nuclear symmetry energy that opposes such a
change~\cite{RocaMaza:2008ja}.  Clearly, both the sequence of
neutron-rich nuclei as well as the neutron-drip density depend
critically on the symmetry energy, which imposes a penalty on the
system as it departs from the symmetric ($N\!=\!Z\!=\!A/2$)
limit. However, nuclear masses with large neutron excess in the region
of the outer crust ($Z\simeq 26-50$) are presently unknown. Thus,
state-of-the-art facilities for rare isotope beams {\sl (FRIB)} that
will map the limits of nuclear existence --- such as the one recently
commissioned at Michigan State University --- will be instrumental in
constraining the structure and composition of the outer crust.

\subsection{The Inner Crust} 

The inner crust of the neutron star comprises the region from
neutron-drip density ($\approx 4\times 10^{11}{\rm g/cm^{3}}$) up to
the density at which uniformity in the system is restored
(approximately $1/3$ to $1/2$ of normal nuclear matter saturation
density). At these densities the system exhibits rich and complex
structures that emerge from a dynamical competition between
short-range nuclear attraction and long-range Coulomb repulsion.  At
the lower densities present in the {\sl outer crust}, these length
scales are well separated and the system organizes itself into a
crystalline lattice of neutron-rich nuclei, while at densities of the
order of half nuclear-matter saturation density, uniformity in the
system is restored. Yet the transition region from the highly-ordered
crystal to the uniform liquid mantle is complex and not well
understood. Length scales that were well separated in both the
crystalline and uniform phases are now comparable, giving rise to {\sl
``Coulomb frustration''}. It has been speculated that the transition
to the uniform phase {\sl must} go through a series of changes in the
dimensionality and topology of these complex structures --- known as
{\sl ``nuclear pasta''}~\cite{Ravenhall:1983uh, Hashimoto:1984}.
 
\begin{figure}[h]
  \includegraphics[height=2.5in]{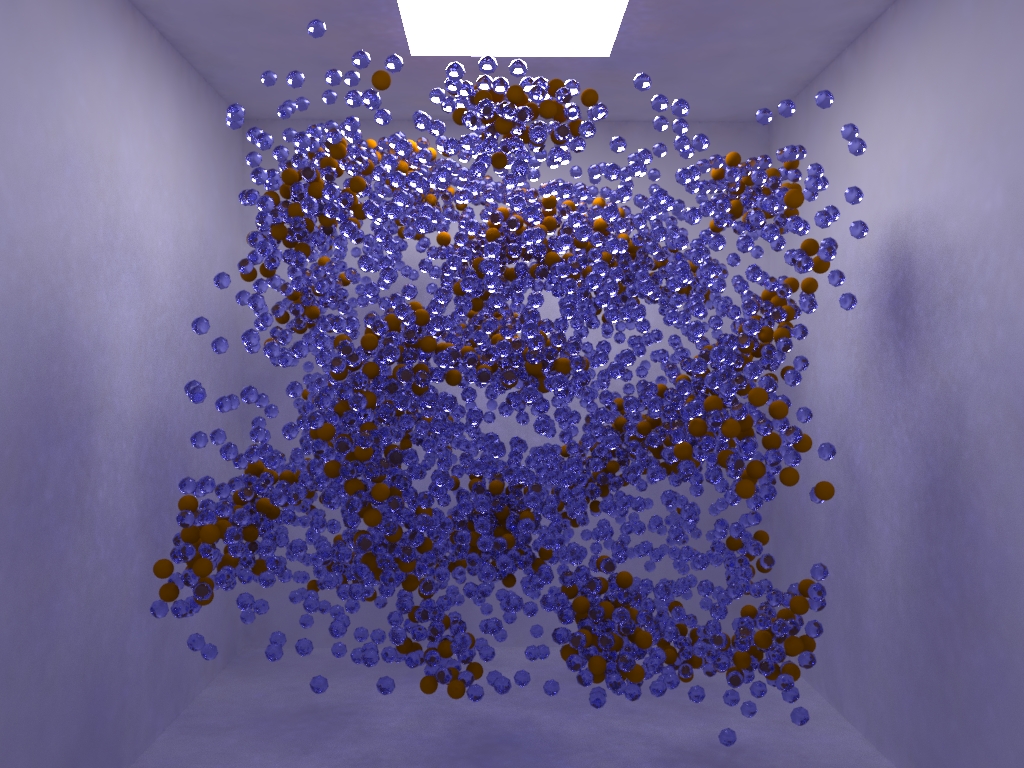}
 \vspace{-0.2cm}
 \caption{A snapshot of a Monte Carlo simulation 
          for a configuration of 4,000 nucleons at a baryon 
          density of $0.025~{\rm fm}^{-3}$ (a sixth of 
          normal nuclear matter saturation density), a proton 
          fraction of $Z/A\!=\!0.2$, and a temperature 
          of $1$~MeV.} 
 \label{Fig2}
\end{figure}
In Fig.~\ref{Fig2} a snapshot obtained from a Monte-Carlo simulation 
of a nuclear system at densities relevant to the inner crust is
displayed~\cite{Horowitz:2004yf,Horowitz:2004pv}. The figure displays
how the system organizes itself into neutron-rich clusters ({\sl
i.e.,} ``nuclei'') of complex topologies that are surrounded by a
vapor of (perhaps superfluid) neutrons. Such complex pasta structures
may have a significant impact on the propagation of neutrinos and
electrons throughout the star.

Interestingly enough, a seemingly unrelated condensed-matter problem
--- that of the strongly-correlated electron gas --- appears to have a
very close connection to the exotic pasta phases. In the case of the
electron gas, one is interested in understanding the transition between
the low-density Wigner crystal and the uniform high-density Fermi
liquid. It has been argued that where a first order phase transition
is expected, intermediate {\sl ``microemulsions''} ({\sl i.e.,}
pasta-like) phases with certain universal characteristics occur
instead.  Indeed, in {\sl two-spatial} dimensions a new theorem
implies that first-order phase transitions are forbidden in the
presence of long-range ({\sl e.g.,} Coulomb)
forces\cite{Jamei:2005}. Note that no generalization of this theorem
exists in three dimensions. Hence, although the existence of pasta
phases in neutron stars may be plausible~\cite{Ravenhall:1983uh,
Hashimoto:1984}, Oyamatsu and Iida have shown in a recent work that
pasta formation may not be universal~\cite{Oyamatsu:2006vd}.  Rather,
the formation (or lack-thereof) of pasta phases is intimately related
to the density dependence of the symmetry energy, a quantity that is
poorly known. In particular, the authors conclude that pasta formation
requires models with a stiff symmetry energy.  Although at present
poorly known, the density dependence of the symmetry energy may be
accurately determined at the Jefferson Laboratory.  The Parity Radius
Experiment (PREx) promises to measure the skin thickness of $^{208}$Pb
accurately and model independently via parity-violating electron
scattering~\cite{Horowitz:1999fk, Michaels:2005}. PREx will provide a
unique experimental constraint on the density dependence of the
symmetry energy due its strong correlation to the neutron skin of
heavy nuclei~\cite{Brown:2000}.

\subsection{The Core} 

Whereas the non-uniform crust displays fascinating and intriguing
dynamics, its structural impact on the star is rather modest.  Indeed,
the crust --- with a thickness of approximately $1$~km --- represents
10\% of the size of the neutron star and contains only about 2\% of its
mass. The range of densities span by the stellar core go from about
$1/2$ up to several times nuclear matter saturation density.  The
highest density attained in the core depends critically on the
equation of state of neutron-rich matter which at those high densities
is poorly constrained. The cleanest constraint on the equation of
state at high-density will emerge as we answer one of the central
questions in nuclear astrophysics: {\sl what is the maximum mass of a
neutron star?} Or equivalently, {\sl what is the minimum mass of a
black hole?}

During the past few years some enlightening correlations between
finite nuclei and several neutron-star properties have been
established~\cite{Horowitz:2000xj,Horowitz:2001ya,
Horowitz:2002mb,Steiner:2004fi}.  This may at first appear surprising
as a heavy nucleus (such as ${}^{208}$Pb) is 18 order of magnitudes
smaller and 55 orders of magnitude lighter than a neutron star. Yet
remarkably, both the neutron radius of a heavy nucleus and the radius
of a neutron star depend critically on the pressure of neutron-rich
matter. Such correlation among objects of such a disparate size is not
difficult to understand. Heavy nuclei develop a neutron-rich skin as a
result of both a large neutron excess and a Coulomb barrier that
reduces the proton density at the surface of the nucleus. The neutron
skin depends critically on the pressure (which is related to the
derivative of the energy with respect to the density) exerted on the
surface neutrons. It is precisely this same pressure that supports a
neutron star against gravitational collapse. Thus models that predict
larger neutron skins in heavy nuclei also tend to predict larger
neutron star radii. This correlation is displayed in Fig.~\ref{Fig4}.
In this figure two families of accurately calibrated relativistic mean
field models (NL3~\cite{Lalazissis:1996rd, Lalazissis:1999} and
``FSUGold''~\cite{Todd-Rutel:2005fa}) are employed to generate a range
of values for the {\sl poorly known} neutron skin in ${}^{208}$Pb.  It
is observed that the stiffer the equation of state of neutron-rich
matter, the larger the neutron skin of ${}^{208}$Pb~\cite{Brown:2000}
and the larger the corresponding neutron-star radius. Moreover, the
figure includes a {\sl hypothetical} point that may be determined once
both the neutron radius of a heavy nucleus and the neutron radius 
{\emph {and}} mass of neutron star are accurately determined. If such a
determination reveals a large neutron skin in ${}^{208}$Pb (suggesting
a stiff EOS at low densities) but a small neutron-star radius
(suggesting a ``softening'' of the EOS at high densities) this may be
indicative of a phase transition to an exotic state of matter, such as
meson condensates or quark matter.

\begin{figure}[h]
  \includegraphics[width=4.0in]{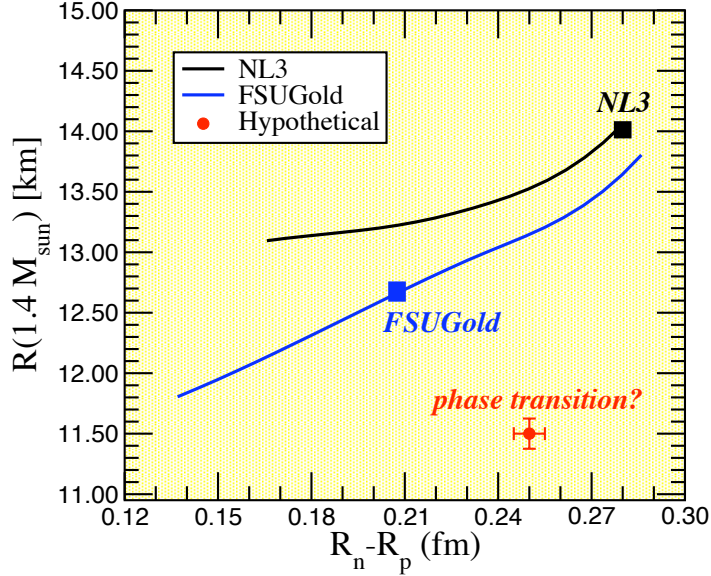}
 \vspace{-0.2cm}
 \caption{The radius of a 1.4 solar-mass neutron star as a 
          function of the neutron skin in ${}^{208}$Pb for
          two families of accurately calibrated relativistic
          mean field models (NL3 and FSUGold). Note that models
          that predict larger neutron skins tend to also 
          predict larger neutron-star radii.}
 \label{Fig3}
\end{figure}

\section{Formalism}


To establish the fundamental role played by the equation of state on
the structure of neutron stars (indeed, on any compact object in
hydrostatic equilibrium) we introduce the Tolman-Oppenheimer-Volkoff
({\sl TOV}) equations, which represent the extension of Newton's laws
to the domain of general relativity. The TOV equations may be expressed 
as a coupled set of first-order differential equations of the following 
form:
\begin{subequations}
 \begin{align}
   & \frac{dP}{dr}=-G\,\frac{{\cal E}(r)M(r)}{r^{2}}
         \left[1+\frac{P(r)}{{\cal E}(r)}\right]
         \left[1+\frac{4\pi r^{3}P(r)}{M(r)}\right]
         \left[1-\frac{2GM(r)}{r}\right]^{-1} \;,
         \label{TOVa}\\
   & \frac{dM}{dr}=4\pi r^{2}{\cal E}(r)\;,
         \label{TOVb}
 \end{align}
 \label{TOV}
\end{subequations}
where $G$ is Newton's gravitational constant, while $P(r)$, ${\cal
E}(r)$, and $M(r)$ represent the pressure, energy density, and
enclosed-mass profiles of the star, respectively. The last three terms
(enclosed in square brackets) in Eq.~(\ref{TOVa}) represent the
extension of Newton's laws to the relativistic domain. To the extent
that neutron stars may be regarded as spherically symmetric objects in
hydrostatic equilibrium, the use of the TOV equations hinges
exclusively on the validity of general relativity, an extremely safe
assumption.  Remarkably then, the only input that neutron stars are
sensitive to is the equation of state of neutron-rich matter. This may
be easily appreciated as the changes in pressure and enclosed mass as
a function of radius (left-hand side of the equations) depend not only
on the values of these quantities at $r$, but also on the ``unknown''
energy density ${\cal E}(r)$ of the system. Thus, no solution of the
TOV equations is possible until an equation of state (namely, a $P$
{\sl vs} ${\cal E}$ relation) is supplied.


Relativistic mean-field descriptions of the ground-state properties of
medium to heavy nuclei have enjoyed enormous success. These highly
economical descriptions encode a great amount of physics in a handful
of model parameters that are calibrated to a few ground-state
properties of a representative set of medium to heavy nuclei. An
example of such a successful paradigm is the relativistic NL3
parameter set of Lalazissis, Ring, and
collaborators~\cite{Lalazissis:1996rd,Lalazissis:1999}.

The Lagrangian density employed in this work is rooted on the seminal
work of Walecka, Serot, and their many collaborators (see
Refs.~\cite{Walecka:1974qa,Serot:1984ey,Serot:1997xg} and references
therein). Since first published by Walecka more than three decades
ago~\cite{Walecka:1974qa}, several refinements have been implemented
to improve the quantitative standing of the model. In the present work
we employ an interacting Lagrangian density of the following
form~\cite{Horowitz:2000xj,Todd-Rutel:2005fa,Mueller:1996pm}:
\begin{align}
{\mathscr L}_{\rm int} & =
 \bar\psi\left[g_{\rm s}\phi   \!-\!
         \left(g_{\rm v}V_\mu  \!+\!
    \frac{g_{\rho}}{2}\tau\cdot{\bf b}_{\mu}
                               \!+\!
    \frac{e}{2}(1\!+\!\tau_{3})A_{\mu}\right)\gamma^{\mu}
         \right]\psi \nonumber \\
                   & -
    \frac{\kappa}{3!} (g_{\rm s}\phi)^3 \!-\!
    \frac{\lambda}{4!}(g_{\rm s}\phi)^4 \!+\!
    \frac{\zeta}{4!}
    \Big(g_{\rm v}^2 V_{\mu}V^\mu\Big)^2 \!+\!
    \Lambda_{\rm v}
    \Big(g_{\rho}^{2}\,{\bf b}_{\mu}\cdot{\bf b}^{\mu}\Big)
    \Big(g_{\rm v}^2V_{\mu}V^\mu\Big) \;.
 \label{Lagrangian}
\end{align}
The original Lagrangian density of Walecka consisted of an isodoublet
nucleon field ($\psi$) together with neutral scalar ($\phi$) and
vector ($V^{\mu}$) fields coupled to the scalar density
($\bar\psi\psi$) and conserved nucleon current
($\bar\psi\gamma^{\mu}\psi$), respectively~\cite{Walecka:1974qa}. In
spite of its simplicity (indeed, the model contains only two
dimensionless coupling constants), symmetric nuclear matter saturates
even when the model was solved at the mean-field
level~\cite{Walecka:1974qa}. By adding additional contributions from a
single isovector meson (${\bf b}^{\mu}$) and the photon ($A^{\mu}$),
Horowitz and Serot~\cite{Horowitz:1981xw} obtained results for the
ground-state properties of finite nuclei that rivaled some of the most
sophisticated non-relativistic calculations of the time. However,
whereas the two dimensionless parameters in the original Walecka model
could be adjusted to reproduce the nuclear saturation point, the
incompressibility coefficient (now a prediction of the model) came out
too large ($K\!>\!500$~MeV) as compared with existing data on
breathing-mode energies~\cite{Youngblood:1977}. To overcome this
problem, Boguta and Bodmer introduced cubic ($\kappa$) and quartic
($\lambda$) scalar meson self-interactions that accounted for a
significant softening of the equation of state
($K\!=\!150\!\pm\!50$~MeV)~\cite{Boguta:1977xi}.  

\section{Results}

\begin{figure}[h]
  \includegraphics[width=4.5in]{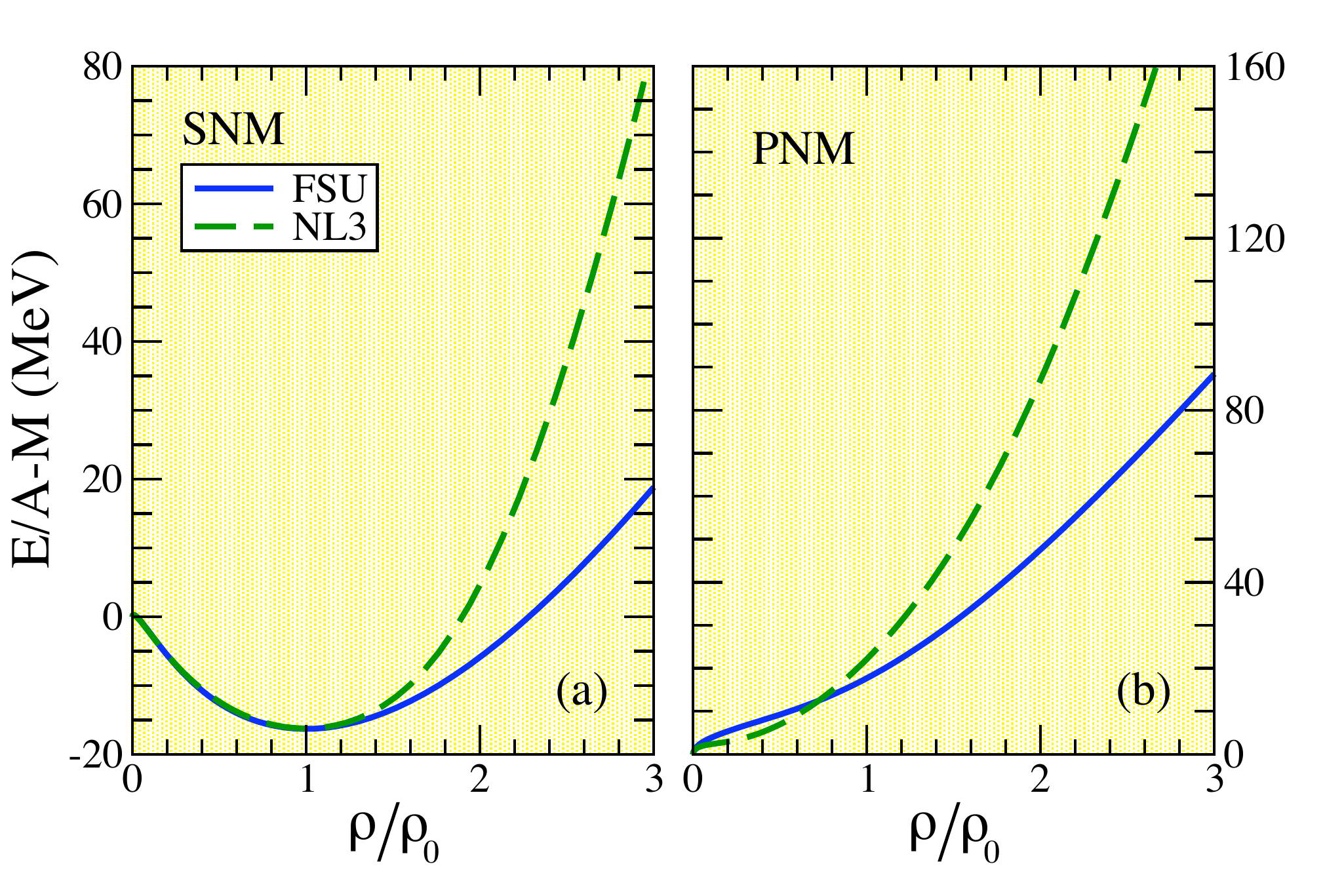}
 \vspace{-0.2cm}
 \caption{Equation of state (energy per particle as a function
          of baryon density) for symmetric matter (a) and pure
          neutron matter (b) as predicted by the FSUGold (blue 
          solid line) and NL3 (green dashed line) models.}
 \label{Fig4}
\end{figure}

The two parameters of the Lagrangian density that remain to be
discussed are $\zeta$ and $\Lambda_{\rm v}$. To underscore their
fundamental role we now proceed to compute the equation of state of
nuclear matter (namely, the energy per particle as a function of
density and neutron excess) and its impact on the {\sl Mass vs Radius}
relationship of neutron stars. To capture the behavior of the EOS with
neutron excess we display in Fig.~\ref{Fig4} the EOS for the two
extreme cases of symmetric-nuclear and pure-neutron matter. From the
EOS the various bulk parameters that characterize the behavior of
neutron-rich matter around saturation density may be extracted; these
parameters are listed in Table~\ref{Table1}.  For a more detailed
discussion of the evolution of the EOS with neutron excess, including
a precise definition of the various bulk parameters, see
Ref.~\cite{Piekarewicz:2008nh}.  Note that the symmetry energy is
given to a very good approximation as the difference between the
energy-per-particle of pure neutron matter and that of symmetric
nuclear matter. As we have done elsewhere~\cite{Piekarewicz:2007us},
our results were generated using two accurately calibrated models:
NL3~\cite{Lalazissis:1996rd,Lalazissis:1999} and
FSUGold~\cite{Todd-Rutel:2005fa}.
 
Whereas the FSUGold and NL3 models agree on the energy and density at
saturation, quantities that are tightly constrained by existent
ground-state observables, significant discrepancies emerge in all
remaining parameters. The main difference between the two models may
be succinctly summarized by stating that whereas FSUGold predicts a
soft behavior for both symmetric nuclear matter (through $K_{0}$) and
the symmetry energy (through $L$), NL3 predicts a stiff behavior for
both. Note that {\sl ``stiff''} or {\sl ``soft''} refers to whether
the energy increases rapidly or slowly with density.

\begin{table}
\begin{tabular}{|l||c|c|c|c|c|c|c|c|}
 \hline
 Model & $\rho_{0}$ & $\varepsilon_{0}$
       & $K_{0}$ & $Q_{0}$ & $J$ & $L$ 
       & $K_{\rm sym}$  & $\gamma$\\
 \hline
 FSU     &  0.148   & $-$16.30 & 230.0 & $-$523.4 & 32.59 &  60.5
         & $-$51.3  & 0.64 \\
 NL3     &  0.148   & $-$16.24 & 271.5 & $+$204.2 & 37.29 & 118.2
         & $+$100.9 & 0.98 \\
\hline
\end{tabular}
\caption{Bulk parameters characterizing the energy of symmetric 
         nuclear matter and the symmetry energy at saturation 
         density. All quantities are in MeV, with the exception 
         of $\rho_{0}$ given in fm$^{-3}$ and the dimensionless 
         parameter $\gamma$. For a detailed explanation of all
         these quantities see Ref.~\cite{Piekarewicz:2008nh}.} 
\label{Table1}
\end{table}

\begin{figure}[h]
  \includegraphics[width=4.0in]{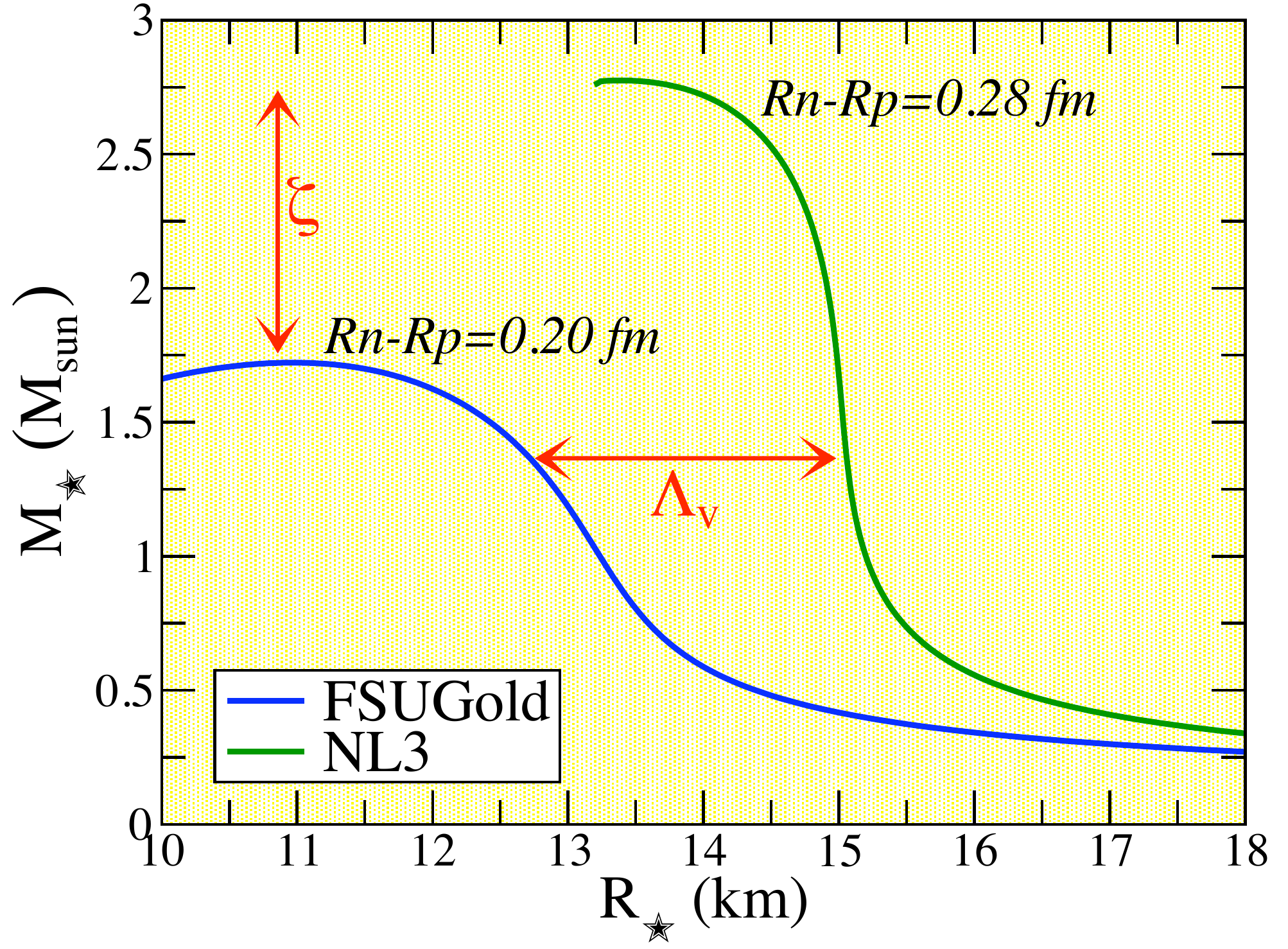}
 \vspace{-0.2cm}
 \caption{Predictions for the {\sl Mass vs Radius} relationship 
          of a neutron star according to the FSUGold (blue solid 
          line) and NL3 (green dashed line) models.}
 \label{Fig5}
\end{figure}

It is worth noting that in the enormously successful NL3 model
both parameters $\zeta$ and $\Lambda_{\rm v}$ are set to zero,
suggesting that the experimental data used in the calibration
procedure is insensitive to the physics encoded in these
parameters. Indeed, M\"uller and Serot found possible to build models
with different values of $\zeta$ that reproduce the same observed
properties at normal nuclear densities, but which yield maximum
neutron star masses that differ by almost one solar
mass~\cite{Mueller:1996pm}. This result indicates that observations of
massive neutron stars --- rather than laboratory experiments --- may
provide the only meaningful constraint on the high-density component
of the equation of state. Finally, the isoscalar-isovector coupling
constant $\Lambda_{\rm v}$ was added in Ref.~\cite{Horowitz:2000xj} to
soften the density dependence of the symmetry energy, which is found 
to be traditionally hard in relativistic mean-field models. Note that
this softening was accomplished without compromising the success of
the model in reproducing ground-state observables that are accurately
measured, such as nuclear masses and charge radii. The
introduction of $\Lambda_{\rm v}$ yields a significant softening of 
the density dependence of the symmetry energy and, correspondingly,
significant reductions in both the neutron skin of heavy nuclei and
the radii of neutron stars~\cite{Horowitz:2001ya}. To illustrate this
behavior we display in Fig.~\ref{Fig5} the {\sl Mass vs Radius}
relationship of a neutron star. The curves are labeled by their
corresponding neutron-skin thickness in ${}^{208}$Pb. Note that the
reduction in the value predicted by FSUGold relative to NL3 is due to
the inclusion of $\Lambda_{\rm v}$; so is the reduction in
neutron-star radii. For example, for a 1.4 solar-mass neutron star the
reduction in the neutron-star radius is close to 2 km. As previously
asserted, this correlation suggests how the measurement of the neutron
radius of a single heavy nucleus --- as will be done at the Jefferson
Laboratory in the case of ${}^{208}$Pb --- will constrain the radii of
neutron stars. In contrast, the reduction in the maximum neutron-star
mass that can be supported by the EOS is controlled by the non-linear
vector coupling $\zeta$. This parameter is at present (and perhaps
forever) {\emph{unconstrained}} by laboratory experiments and its
determination awaits the identification of massive neutron stars.

\section{Conclusions}

The fundamental role played by the equation of state of cold 
neutron-rich matter in determining the properties of neutron
stars was emphasized. Indeed, we observed how the properties
of spherical neutron stars in hydrostatic equilibrium are
only sensitive to the equation of state. As such, neutron 
stars provide a unique window into the behavior of nuclear 
systems under extreme conditions of density and neutron excess.
In this contribution I underscored the special role that nuclear 
physics plays in the determination of the various fascinating
phases displayed in neutron stars and the important role that
a concerted effort by theory, experiment, and observation will
play in their elucidation.

\begin{acknowledgments}
 The author wishes to thank his many collaborators that were 
 involved in this work. This work was supported in part by 
 United States Department of Energy under grant DE-FD05-92ER40750.
\end{acknowledgments}

\vfill\eject
\bibliography{/Users/jorge/Tex/Papers/ReferencesJP}
\end{document}